\documentclass[aps,prl,reprint,superscriptaddress,twocolumn]{revtex4}

\newcommand{\pars}[1]{\left(#1\right)}
\newcommand{\aspas}[1]{``#1"}
\usepackage{amsfonts}
\usepackage{amsmath}
\usepackage[normalem]{ulem}
\usepackage{bm}
\usepackage{graphicx}
\usepackage{braket}
\usepackage{subfig} % for subfigures

\usepackage{float}

\usepackage{subfig}
\usepackage{eqparbox}
\usepackage{amssymb}
\usepackage{lipsum}
\usepackage{caption}
\usepackage{array,multirow}
\usepackage[utf8x]{inputenc}
\usepackage{color}

\begin{document}
	
	\title{
		Swap Test-based Characterization of Quantum Processes in Universal Quantum Computers}

    \author{Pedro Ripper}
	\altaffiliation{Corresponding author: pripper@opto.cetuc.puc-rio.br}
	\affiliation{Center for Telecommunication Studies, Pontifical Catholic University of Rio de Janeiro,\\ 22451-900, Rio de Janeiro, Brazil}

    \author{Gustavo Amaral}
	
	\affiliation{Center for Telecommunication Studies, Pontifical Catholic University of Rio de Janeiro,\\ 22451-900, Rio de Janeiro, Brazil}

    \author{Guilherme Temporão}
	
	\affiliation{Center for Telecommunication Studies, Pontifical Catholic University of Rio de Janeiro,\\ 22451-900, Rio de Janeiro, Brazil}

	\begin{abstract}
		Quantum Computing has been presenting major developments in the last few years,  unveiling systems with a increasing number of qubits. However, unreliable quantum processes in universal quantum computers still represent one of the the greatest challenges to be overcome. Such obstacle has its source on noisy operations and interactions with the environment which introduce decoherence to a quantum system. In this article we verify whether a tool called Swap Test is able to identify decoherence. Our experimental results demonstrate that the Swap Test can be employed as an alternative to a full Quantum Process Tomography, with the advantage of not destroying the qubit under test, under certain circumstances, as long as some modifications are introduced. 
		\end{abstract}
	
	\maketitle

%%%%%%%%%%%%%%%%%%%%%%%%%%%%%%

\section{Introduction}
\label{intro}
Quantum Computing has been gaining traction among other technologies that will revolutionize our future. The expectation that quantum computers will harness what is known as \textit{Quantum Speedup} has recently placed this field under the spotlight, which attracted huge investments from both governmental and private sectors. Yet, we are in a stage of Quantum Computing called NISQ era\citet{Preskill_2018}, that is characterized by quantum computers that are still very vulnerable to unreliable operations and decoherence by interactions between its system and the environment. Much has been done in the field for characterizing quantum computers' performance amid these shortcomings, such as Randomized Benchmarking\citet{helsen2020general} for gate fidelity estimation and Quantum Volume\citet{PhysRevA.100.032328} evaluation for a more general metric about a machine's capacity.

Quantum decoherence has also been a topic of study outside the context of Quantum Computing. In the framework of photonics, for example, said phenomenon can be observed as depolarization. In a previous work by our group \citet{amaral2018complementarity}, we have shown how two-photon interference in a Hong-Ou-Mandel (HOM) interferometer \citet{PhysRevLett.59.2044} can be exploited in order to determine whether decoherence has taken place. However, the HOM phenomenon is an effect restricted to the context of photonics. Aiming for a device-agnostic framework, we employ the Swap Test (ST)\citet{Buhrman_2001} as an alternative, due to the fact that it has been proven equivalent to the HOM effect for input pure states\citet{garcia2013swap}.

\begin{figure}[ht]
    \centering
  \includegraphics[width=0.9\linewidth]{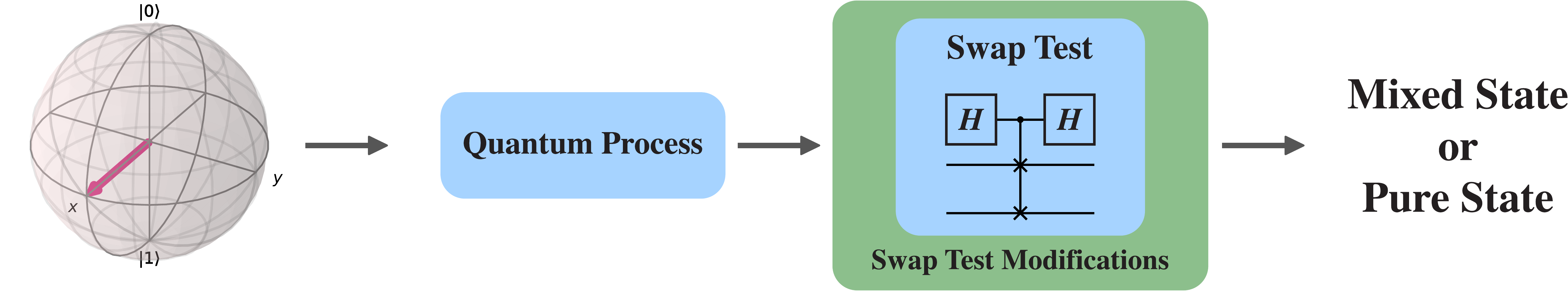}
\caption{Motivation: using the Swap Test to check for a possible decoherence process that an unknown qubit could have undergone. The output of the process should identify the input state as a mixed or pure state.}
\label{fig:intro}       % Give a unique label
\end{figure}

Here, we formalize a previously nonexistent ST-based methodology to characterize an unknown quantum process or an unknown qubit with varying purity from 0 (completely mixed) to 1 (pure). We also present the outcomes from our research from previous works about the ST\citet{Buhrman_2001,garcia2013swap,Cincio_2018,foulds2021controlled} followed by shortcomings of this technique. Yet, as will be shown in this work, the standard ST setting is not able to correctly characterize a qubit in a mixed state. Therefore, we propose adjustments to the current ST protocol, in order to check whether a qubit state has been affected by decoherence or, rather, a random unitary operation - see Figure \ref{fig:intro}. All the herewith presented experiments were conducted using IBM's Qiskit\citet{cite_qiskit} framework to simulate quantum circuits and execute them on real quantum computers.

\section{The Swap Test}

The impossibility of deterministically distinguishing between two general unknown quantum states is one of the core features of quantum mechanics \citet{aperes}. Therefore, methods that provide information on how close two quantum states $\ket{\phi}$ and $\ket{\psi}$ are from one another - such as the \textit{fidelity} $F = |\braket{\psi | \phi}|^2$ - must be probabilistic. For example, one can verify whether two qubits are identical to each other by projecting their joint state onto the singlet two-qubit state $\ket{\psi^-} = \tfrac{1}{\sqrt{2}}(\ket{0}\ket{1}-\ket{1}\ket{0})$, where $\{\ket{0},\ket{1}\}$ represent the computational basis elements of a single qubit Hilbert Space. Indeed, it is easy to show that

\begin{equation} \label{eq:psiminusprojection}
\bra{\psi^-} [ \ket{\phi}\ket{\psi}] = 0 \iff \ket{\psi} = \ket{\phi}
\end{equation}
which means that a full Bell-State Measurement (BSM) has zero probability of producing an outcome associated to the singlet state if the two qubits are identical. Therefore, by repeating the experiment multiple times, it is possible to determine, up to an arbitrarily high degree of confidence, whether two input states are identical or not. It is important to note that even a single measurement can provide information: if the projection onto the singlet state is successful, than one can predict with certainty that the input states have fidelity strictly less than one.

However, the projection onto the singlet state has some limitations; one of them is that it is a destructive procedure. Even if the two input states are identical, they are destroyed in the process. Introduced by Buhrman \textit{et al} in the setting of Quantum Fingerprinting\citet{Buhrman_2001}, the \textit{Swap Test} is a class of methods that can also distinguish between two quantum states $\ket{\phi}$ and $\ket{\psi}$, with the advantage of being able to not destroy the states (by employing an ancilla qubit). There are different variations of quantum circuits that are capable of realising the operation known as the Swap Test. In our work we focused on two, that we shall call the Controlled-Swap (CSWAP) Swap Test, which is of the nondestructive kind, and the Bell State Measurement(BSM) Swap Test, which is very close to the projection onto the singlet state shown above.

% \subsection{The CSWAP Swap Test}

\begin{figure}[ht]
\centering
  \includegraphics[scale=0.9]{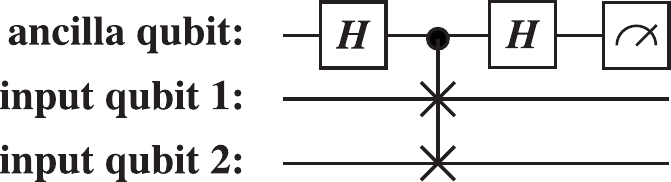}
\caption{Controlled-Swap (CSWAP) ST circuit.}
\label{fig:1}       % Give a unique label
\end{figure}

The CSWAP ST is the original ST setting, which was presented by Buhrman \textit{et al}\citet{Buhrman_2001}. It  is comprised of three qubits. The first one is an ancillary qubit in the $ \ket{0}$ state. The other qubits, $\ket{\phi}$ and $\ket{\psi}$, are the ones which will have their states compared. In the circuit, firstly, a Hadamard operator is applied on the ancillary qubit. In the Controlled-Swap gate that follows, the ancilla acts as a control qubit, whereas the Swap gate targets the remaining states. Another Hadamard gate succeeds the operation, carrying out the transformation on the ancillary qubit. The CSWAP ST circuit is portrayed in Figure \ref{fig:1} and its operations unfold as shown in Equation \ref{eq1}.

 \begin{gather}
    \ket{0}\ket{\psi}\ket{\phi} \xrightarrow[]{H} \frac{\ket{0} + \ket{1}}{\sqrt{2}}\ket{\psi}\ket{\phi} \nonumber 
      \xrightarrow[]{CSWAP} \\
      \frac{\ket{0}\ket{\psi}\ket{\phi} + \ket{1}\ket{\phi}\ket{\psi}}{\sqrt{2}} \nonumber 
        \xrightarrow[]{H} 
        \frac{\frac{\ket{0} + \ket{1}}{\sqrt{2}}\ket{\psi}\ket{\phi} + \frac{\ket{0} - \ket{1}}{\sqrt{2}}\ket{\phi}\ket{\psi}}{\sqrt{2}} 
           \\ \xrightarrow[]{}   \frac{\ket{0}(\ket{\psi}\ket{\phi}+\ket{\phi}\ket{\psi}) + \ket{1}(\ket{\psi}\ket{\phi}-\ket{\phi}\ket{\psi})}{2} 
    \label{eq1}    
\end{gather}

Following the operations, a canonical basis measurement is performed on the ancillary qubit, and according to Equation \ref{eq2} , two outcomes are possible. First, if states $\ket{\phi}$ and $\ket{\psi}$ are indistinguishable, $\ket{0}$ is found with a 100\% probability. On the other hand, if they are distinct, both $\ket{0}$ and $\ket{1}$ are possible outcomes, with a specific probability which can be estimated by referencing Equation \ref{eq2}.

\begin{gather}
P(\ket{0}) = \frac{(\bra{\psi}\bra{\phi}+\bra{\phi}\bra{\psi})(\ket{\psi}\ket{\phi}+\ket{\phi}\ket{\psi})}{4}\nonumber \\
P(\ket{0}) =\frac{2\braket{\psi|\phi}^2 + 2}{4}
=\frac{\braket{\psi|\phi}^2 + 1}{2} \nonumber \\ 
P(\ket{1}) = 1-P(\ket{0})= \frac{1-\braket{\psi|\phi}^2}{2}
\label{eq2}
\end{gather}

Consequently, in order to find whether two states are equal, it would be necessary to realize several iterations of the ST with no occurrences of the outcome 1. On the other hand, a single result 1 from the ST would be sufficient to prove that the states are distinguishable. Such circumstances show that a single outcome 0 from the ST does not give any real information about the input qubits, leaving the ST attempt inconclusive.

% \subsection{The Bell State Measurement Swap Test}

The CSWAP ST circuit, previously presented in Figure \ref{fig:1}, has been offered some modifications on different articles\citet{Cincio_2018,garcia2013swap}, being the BSM ST one of them. Although there were no changes in the probability distribution of outcomes between the CSWAP and BSM ST, the BSM ST arrangement provides a significant modification: the BSM ST does not need an ancilla qubit as the CSWAP ST, it is comprised of a Bell State Measurement on the input qubits, performed by a CNOT gate followed by a Hadamard gate. Having said that, the ancilla measurement is substituted by a post-measurement classical processing. The final result from the BSM ST is 1 if both qubits are measured as a state $\ket{1}$. Any of the other possible measurements yields 0 for the final outcome. Despite of the lack of an ancilla and the post-measuring processing, the BSM ST probabilities of output follow the ones from the CSWAP ST.

It should be pointed out that the usage of the ST for fidelity estimation was predicted for qubits in \textit{pure} states. With that in mind, in this paper we present a ST attempt for a qubit in a mixed state with another in a pure state. As will be seen, the ST is unable to perform such comparison and requires an adaptation, which will be presented in Section \ref{stCtrlMeasSec}, that makes possible to identify whether a qubit is in a mixed state.

% % For one-column wide figures use
% \begin{figure}[ht]
% \centering
% % Use the relevant command to insert your figure file.
% % For example, with the graphicx package use
%   \includegraphics[scale=1.0]{bsm_st_circuit.pdf}
% % figure caption is below the figure
% \caption{Bell State Measurement Swap Test circuit}
% \label{fig:2}       % Give a unique label
% \end{figure}

\subsection{Characterization Methodology with the Swap Test}

As our main goal was to obtain more information about a qubit in a "unknown" state by detecting whether it has suffered decoherence or not, we established that it would be used as an input for the ST alongside another qubit in a "known" (pure) state. On that premise, our known and unknown qubits are identified as the \textit{control} and \textit{test} qubits respectively. This identification was used in all of our experiments. Moreover, we set a protocol to run our experiments. First, in order to prepare our \textit{control} and \textit{test} qubits, we inserted a rotation operation for the \textit{test} qubit called $R_{prep}$, which could be realized over any axis and a more general unitary operation $U_{ctrl}$ for the \textit{control} qubit initialization. 

\begin{figure}[ht]
  \centering
  \includegraphics[width=1.0\linewidth]{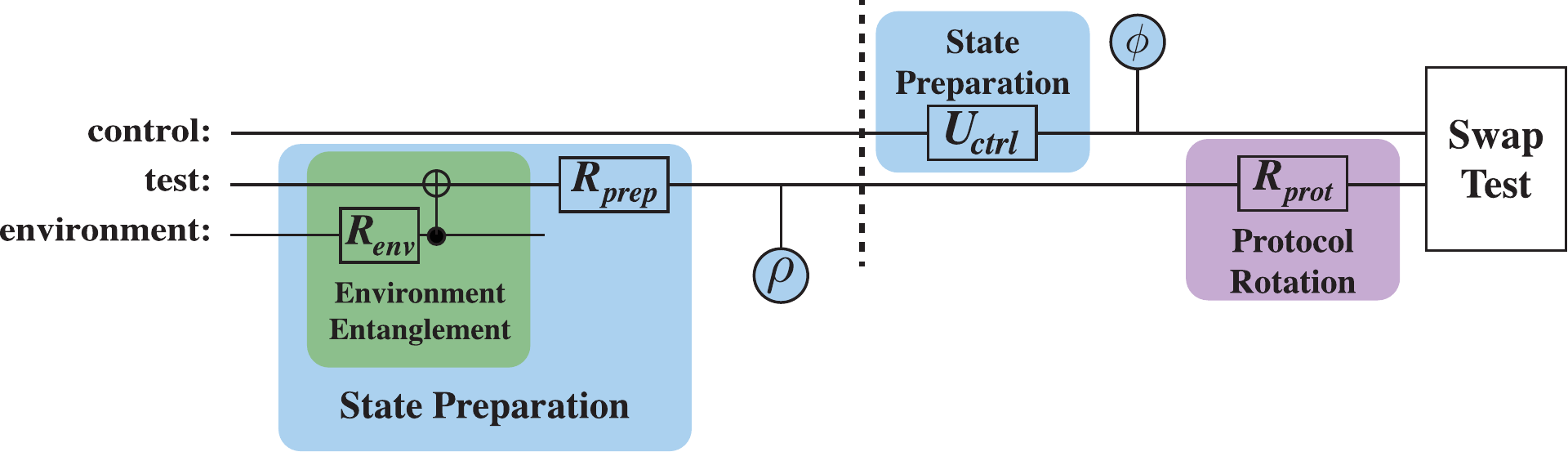}
  \caption{Configuration for Swap Test Experiments.}
  \label{fig:protocolScheme}       % Give a unique label
\end{figure}

Furthermore, as we were interested in running tests with a decoherence process, we added a new auxiliary qubit called \textit{environment} that will have a varying degree of entanglement with the \textit{test} qubit dictated by a rotation along the Y-Axis, that we called $R_{env}$. As the \textit{environment} qubit is initialized at the $\ket{0}$ state, it goes from no-entanglement to full-entanglement with the \textit{test} qubit when a rotation of $\frac{\pi}{2}$ radians over the Y-Axis is made, which leaves its state equal to $\ket{+}$. As consequence, the \textit{test} qubit's \textit{state purity} - which we define as the norm of the corresponding Bloch vector - tends to decrease. Also, we added a rotation operation named $R_{prot}$, which could be performed along any axis for the \textit{test} qubit after its initialization, for evaluating its state evolution to the ST results. Said operations and the state preparation for the \textit{test} and \textit{control} qubits are presented in Figure \ref{fig:protocolScheme}. 

From left to right, we have the following. The \aspas{test} qubit, after the $R_{prep}$ operation, is an unknown state $\rho$ we wish to characterize, which can be coupled to an auxiliary qubit called \aspas{environment} in order to simulate decoherence. The rotation dictated by the unitary operator $R_{env}$ determines the purity of the resulting mixed state (represented by $\rho$) when a partial trace over the environment is performed. An arbitrary rotation $R_{prot}$ ensures that the input state can be anywhere in the interior of the Bloch sphere. The Swap Test is then applied to this state and a known qubit \aspas{control} (which, even though is a pure state, is represented by a density matrix $\sigma$), initialized with an unitary operation $U_{ctrl}$.

\subsection{Swap Test Decoherence Characterization Challenges}

In our attempt to characterize a decoherence process using the ST, we performed several ST iterations varying the $R_{env}$ and $R_{prot}$. For each $R_{env}$ value, that ranged from $0$ to $\frac{\pi}{2}$ radians, we increased $R_{prot}$ from $0$ to $\pi$ radians. As it is expected, the \textit{test} and  \textit{environment} qubits become more entangled as $R_{env}$ gets closer to $\frac{\pi}{2}$ radians and with that, the \textit{test} qubit loses its purity, as seen in Figure \ref{fig:test_decoherence_{env}v}. This experiment was tested for the \textit{test} qubit being $\ket{0}$, $\ket{+i}$ or $\ket{+}$ and with the \textit{control} being set always as $\ket{+}$ or following the \textit{test} initial state. Also, we analysed the differences when the $R_{prot}$ operation was on the Y-Axis or X-Axis. 

\begin{figure}[ht]
    \centering
  \includegraphics[width=0.8\linewidth]{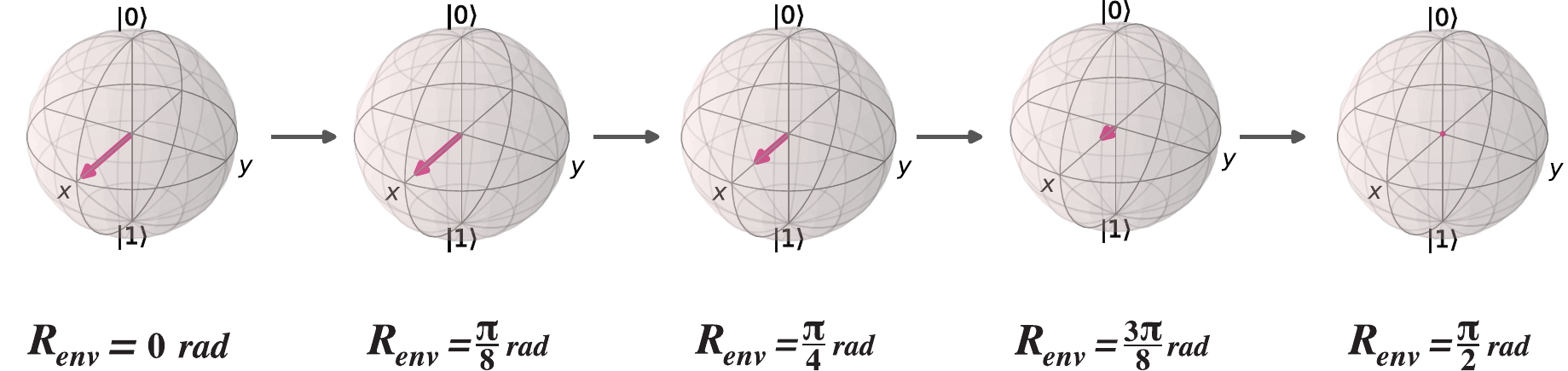}
\caption{Bloch sphere representation of the \textit{test} qubit after the CNOT operation according to $R_{env}$, when it is initially set to the state $\ket{+}$. The purity of the state varies from 1 to 0.}
\label{fig:test_decoherence_{env}v}       % Give a unique label
\end{figure}

\begin{figure}[ht]
  \centering
  \includegraphics[width=1.0\linewidth]{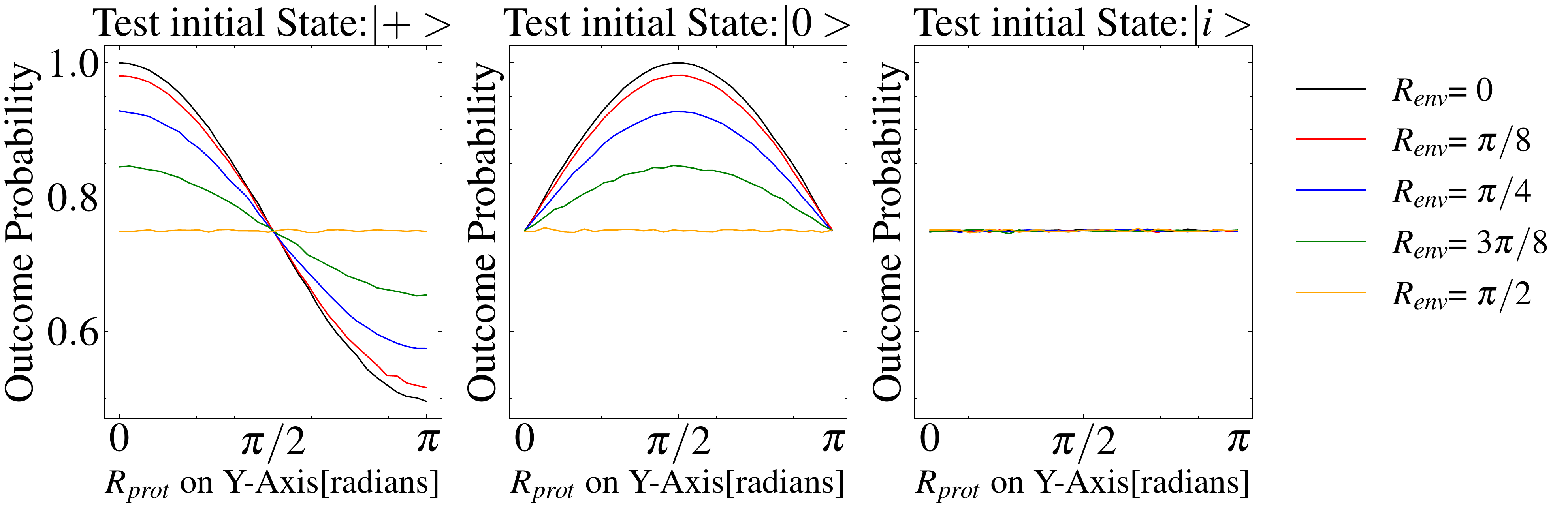}
  \caption{Decoherence Characterization iteration using the Swap Test with \textit{control} qubit set to $\ket{+}$ and $R_{prot}$ set on the Y-Axis. In this example, the CSWAP configuration was used; similar results are obtained for the BSM variation.}
  \label{fig:protocol_original_cswap_ctrlH_ry}       % Give a unique label
\end{figure}

\begin{figure}[ht]
  \centering
  \includegraphics[width=1.0\linewidth]{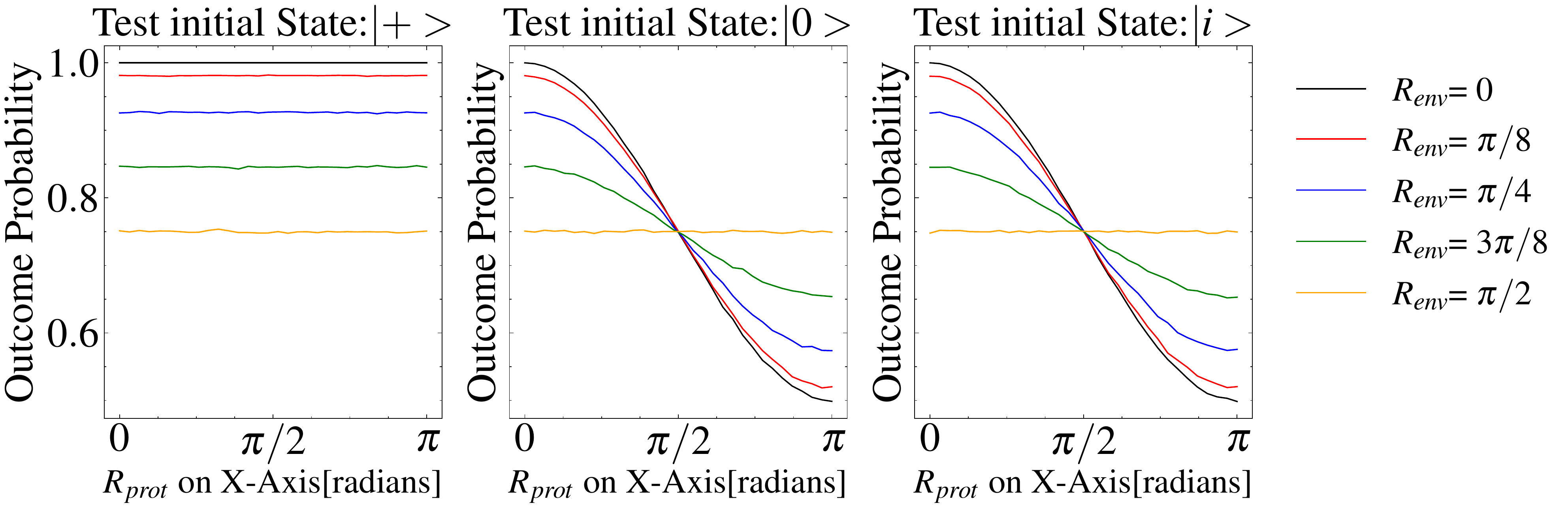}
  \caption{Decoherence Characterization iteration using the Swap Test with \textit{control} qubit's state equals to \textit{test} qubit's state and $R_{prot}$ set on the X-Axis. In this example, the BSM configuration was used; similar results are obtained for the CSWAP variation.}
  \label{fig:protocol_original_bsm_ctrlT_rx}       % Give a unique label
\end{figure}

Unfortunately, our experiments show that the results of the ST are not unique for each input qubit, regardless of the chosen ST variant (CSWAP or BSM). First, the purity of the \textit{test} state did not influence the probability of obtaining output \aspas{1} in the ST whenever it was initialized as $\ket{+i}$, as seen in Figure \ref{fig:protocol_original_cswap_ctrlH_ry}. This means that any state lying along the Y-axis in Bloch sphere would produce the same result. As the rotation is being performed precisely around the Y-axis, one may be tempted to say that a change of axis would suffice to eliminate this redundancy; but it is not the case, as presented in Figure \ref{fig:protocol_original_bsm_ctrlT_rx}, which now performs rotations along the X-axis. In this new scenario, results from initial states $\ket{0}$ and $\ket{1}$ are now indistinguishable. Similar results are obtained when rotating along the Z-axis. It should be mentioned that the results shown in Figs. \ref{fig:protocol_original_cswap_ctrlH_ry} and \ref{fig:protocol_original_bsm_ctrlT_rx} are very similar in both ST variations (CSWAP and BSM). With that said, the ST currently cannot be used to characterize a decoherence process and distinguish its results for different states.

\section{Swap Test with Control Measurement}
\label{stCtrlMeasSec}
\begin{figure}[ht]

  \centering

\subfloat{{\includegraphics[width=0.3\textwidth]{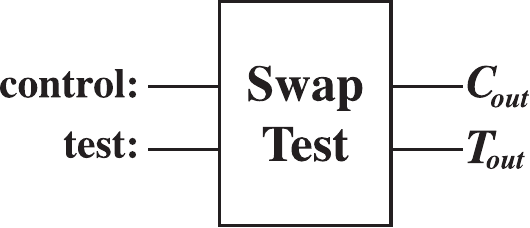}}}

\subfloat{{\includegraphics[width=0.3\textwidth]{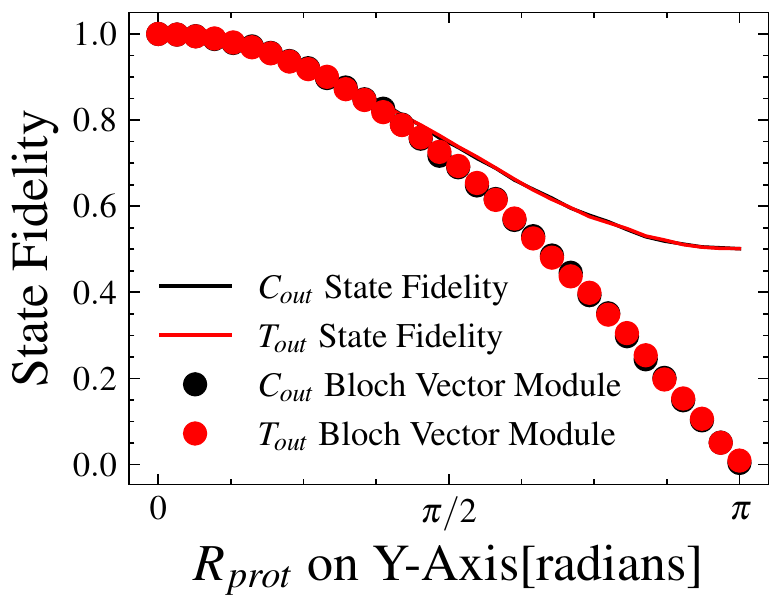}}}

% \begin{subfigure}{0.5\linewidth}
%   \centering
% \includegraphics[width=\linewidth]{st_out_scheme.pdf}

% %   \caption{Separated Outcomes}

% \end{subfigure}%
\vspace{1em}

% \begin{subfigure}{0.7\linewidth}
%   \centering
%   \includegraphics[width=\linewidth]{std_st_state_fidelity.pdf}
% %   \caption{Control and Test's }
% \end{subfigure}%

\caption{State Fidelity and State Purity (Bloch vector norm) from Control and Test's states after undergoing the Swap Test ($C_{out}$ and $T_{out}$). Fidelities are with respect to the corresponding inputs.}
\label{fig:StateFidelityST}

\end{figure}

Amid these shortcomings from the ST, we proposed an adaptation to our initial ST protocol (Figure \ref{fig:protocolScheme}). As our main goal was to characterize a decoherence process on a single qubit which we called \textit{test} by performing the ST with another qubit labelled \textit{control}, we decided to obtain the \textit{control} measurement in parallel with the ST outcome, leaving the \textit{test} qubit unobserved. We aimed to extract more information from the \textit{test} qubit by analysing the ST result along with the \textit{control} qubit measurement outcome. This decision was made owning to the fact that the ST conserves both \textit{test} and \textit{control} qubits if their states overlap. Such fact has been highlighted in \citet{foulds2021controlled} and can be seen in Figure \ref{fig:StateFidelityST} where we calculated the state fidelity of both qubits after the ST($C_{out}$ and $T_{out}$, corresponding to control and test, respectively) in comparison with their initial state. As seen, as the \textit{control} and \textit{test} qubits' states become closer to being orthogonal, they start losing fidelity to their initial state before the ST.


\begin{figure}[ht]
\centering
  \includegraphics[width=1.0\linewidth]{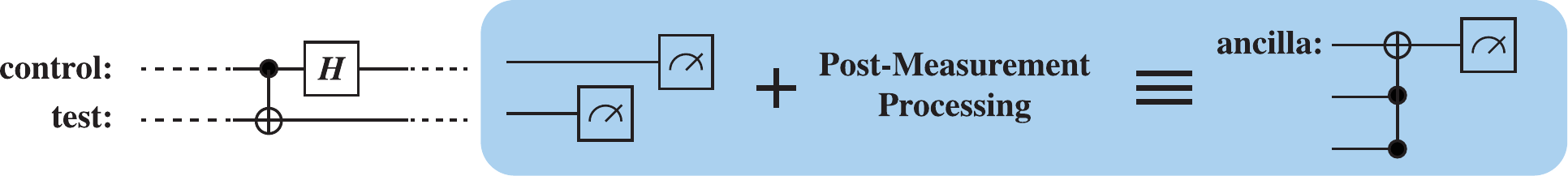}
\caption{BSM ST translation into the Toffoli ST}
\label{fig:BSMtoToffoli}       % Give a unique label
\end{figure}

The State Fidelity\citet{10.5555/1972505} is a metric that shows how close two quantum states are. Those states should be represented as density matrices and their fidelity can be found by Equation \ref{eq:fidelity}. As seen, in order to calculate their fidelity, the density matrix from both states is needed. Being $C_{out}$ and $T_{out}$ unknown, we estimated their density matrices realizing a Quantum State Tomography(QST) procedure \citet{10.5555/1972505}, a method for determining the density matrix of a state given a large number of its copies, represented in Equation \ref{eq:qst}. 
\begin{equation} \label{eq:fidelity}
F(\rho,\sigma) \equiv tr\sqrt{\rho^{1/2}\sigma\rho^{1/2}}
\end{equation}
\begin{equation} \label{eq:qst}
\rho = \frac{I + tr(X\rho)X + tr(Y\rho)Y + tr(Z\rho)Z}{2}
\end{equation}
However, as we have previously seen, the BSM ST requires a measurement on both \textit{test} and \textit{control} qubits. That being said, we decided to use the ST adaptation for the BSM ST previously presented by Cincio \textit{et al}\citet{Cincio_2018} where the measurements along with the post-measurement processing were substituted by a Toffoli gate which had as target an auxiliary qubit, as presented in Figure \ref{fig:BSMtoToffoli}. We referred to this ST modification as the Toffoli ST.

Furthermore, as we would be working with two measurements - on the \textit{control} qubit and on the \textit{ancilla} qubit in both Toffoli and CSWAP Swap Tests - a post-measurement processing was needed in exchange for a single bit output for our experiments. On that premise, observing Figure \ref{fig:StateFidelityST}, if the \textit{control} qubit, which is initially always equal to $\ket{0}$ before the start of any execution, is set to a desired state $\ket{\phi}$ by an unitary operation $U$, its adjoint $U^{\dagger}$ can be added right after the operations for a measurement on the $\ket{\phi}$ basis. From this measurement, it is possible to detect whether the \textit{control} qubit $\ket{\phi}$ was altered during the ST, by being different from \textit{test} qubit, if $Prob(ctrl = 1)>0$. That being said, as an outcome 1 on both \textit{ancilla} or \textit{control} qubit measurements yields distinguishable states, our post-measurement processing is represented by a logic OR operation on both measurements, as presented in Equation \ref{eq:postprocessing}.

\begin{equation} \label{eq:postprocessing}
M = M(\textit{control}) \lor M(ST)
\end{equation}

\subsection{Decoherence Characterization Using the Toffoli Swap Test}

Having our modification for the ST protocol, we tested both  CSWAP and Toffoli Swap Tests with a \textit{control} measurement in a decoherence scheme such as the one in Figure \ref{fig:test_decoherence_{env}v}, in order to check for improvements in the characterization process. Analysing our results, we found one iteration where it was possible to outline a decoherence process without any of the previously reported issues in Figures \ref{fig:protocol_original_cswap_ctrlH_ry} and \ref{fig:protocol_original_bsm_ctrlT_rx}. 

The capability of the Toffoli ST with a \textit{control} measurement to characterize an unknown quantum process can be demonstrated by the following calculations, considering the circuit presented in Figure \ref{fig:protocolScheme}, with the Toffoli ST setting and assuming a rotation $\epsilon$ for $Ry_{env}$, where $0 \leq \theta \leq \frac{\pi}{2}$ and a protocol rotation $\alpha$ on $Rx_{prot}$. Let us consider the initial joint state $\ket{init} = \ket{0}_{env}\ket{0}_{test}\ket{0}_{ctrl}$ representing the environment, test and control qubits. The environment qubit undergoes a rotation $Ry_{env}(\epsilon)$ before a CNOT operation with the test qubit, which was initialized with the $R_{prep} = 0 $, followed by the protocol rotation $Rx_{prot}(\alpha)$ whereas the control qubit undergoes a Hadamard operation ($U_{ctrl} = H$), resulting in the following joint state:

\begin{equation*}
    \begin{split}
    [\cos\pars{\frac{\varepsilon}{2}}\cos\pars{\frac{\alpha}{2}}\ket{0}_{env}\ket{0}_{test} \\ -i\cos\pars{\frac{\varepsilon}{2}}\sin\pars{\frac{\alpha}{2}}\ket{0}_{env}\ket{1}_{test}\\ -i\sin\pars{\frac{\varepsilon}{2}}\sin\pars{\frac{\alpha}{2}}\ket{1}_{env}\ket{0}_{test} \\ +\sin\pars{\frac{\varepsilon}{2}}\cos\pars{\frac{\alpha}{2}}\ket{1}_{env}\ket{1}_{test}] \\ \otimes \pars{\frac{1}{\sqrt{2}}\ket{0}_{ctrl} +\frac{1}{\sqrt{2}}\ket{1}_{ctrl}}
    \end{split}
\end{equation*}
This state can be simplified to:
\begin{equation*}
    \begin{split}
     \frac{1}{2}\cos\pars{\frac{\varepsilon}{2}} \  e^{-i\frac{\alpha}{2}}\ket{0,0,0} + \frac{1}{2}\cos\pars{\frac{\varepsilon}{2}} \  e^{i\frac{\alpha}{2}}\ket{0,0,1} \\ + \frac{1}{2}\cos\pars{\frac{\varepsilon}{2}} \  e^{-i\frac{\alpha}{2}}\ket{0,1,0} - \frac{1}{2}\cos\pars{\frac{\varepsilon}{2}} \  e^{i\frac{\alpha}{2}}\ket{0,1,1} \\ + \frac{1}{2}\sin\pars{\frac{\varepsilon}{2}} \  e^{-i\frac{\alpha}{2}}\ket{1,0,0} - \frac{1}{2}\sin\pars{\frac{\varepsilon}{2}} \  e^{i\frac{\alpha}{2}}\ket{1,0,1} \\ + \frac{1}{2}\sin\pars{\frac{\varepsilon}{2}} \  e^{-i\frac{\alpha}{2}}\ket{1,1,0} + \frac{1}{2}\sin\pars{\frac{\varepsilon}{2}} \  e^{i\frac{\alpha}{2}}\ket{1,1,1}
     \end{split}  
    \end{equation*}
where the notation $\ket{x,y,z}$ was used instead of $\ket{x}_{env}\ket{y}_{test}\ket{z}_{ctrl}$.

Now, we introduce the ancilla qubit and use this state as the input to the Toffoli gate. At the output of the latter, a Hadamard gate is applied to the control qubit rail; we are only interested (due to the OR gate defined in equation 6) in the cases where both control and ancilla are in the $\ket{0}$ state. Therefore, we post-select the components where this is true, and obtain the following (post-selected) output state after a straightforward but somewhat lengthy calculation:
\begin{equation*}
    \begin{split}
     \ket{out}=\frac{1}{\sqrt{2}}\cos\pars{\frac{\varepsilon}{2}} \ \cos\pars{\frac{\alpha}{2}}\ket{0,0,0,0} \\ +  
     \frac{1}{2\sqrt{2}}\cos\pars{\frac{\varepsilon}{2}} \ e^{-i\frac{\alpha}{2}}\ket{0,1,0,0} \\ -
     \frac{1}{\sqrt{2}}\sin\pars{\frac{\varepsilon}{2}} \ \sin\pars{\frac{\alpha}{2}}\ket{1,0,0,0} \\ +
     \frac{1}{2\sqrt{2}}\sin\pars{\frac{\varepsilon}{2}} \ e^{-i\frac{\alpha}{2}}\ket{1,1,0,0},
    \end{split}  
\end{equation*}
where the last qubit is the ancilla qubit. The probability associated to this event can be calculated as: 
\begin{equation*}
    \begin{split}
     Prob(0,0) &= \frac{1}{2}\cos^2\pars{\frac{\varepsilon}{2}}\cos^2\pars{\frac{\alpha}{2}} + \frac{1}{8} \cos^2\pars{\frac{\varepsilon}{2}} \\ & + \frac{1}{2}\sin^2\pars{\frac{\varepsilon}{2}}\sin^2\pars{\frac{\alpha}{2}} +\frac{1}{8} \sin^2\pars{\frac{\varepsilon}{2}} \\ &= \frac{3}{8}+\frac{1}{4}(\cos(\varepsilon)\cos(\alpha)).
    \end{split}  
\end{equation*}

\begin{figure}[ht]
\centering
  \includegraphics[width=\linewidth]{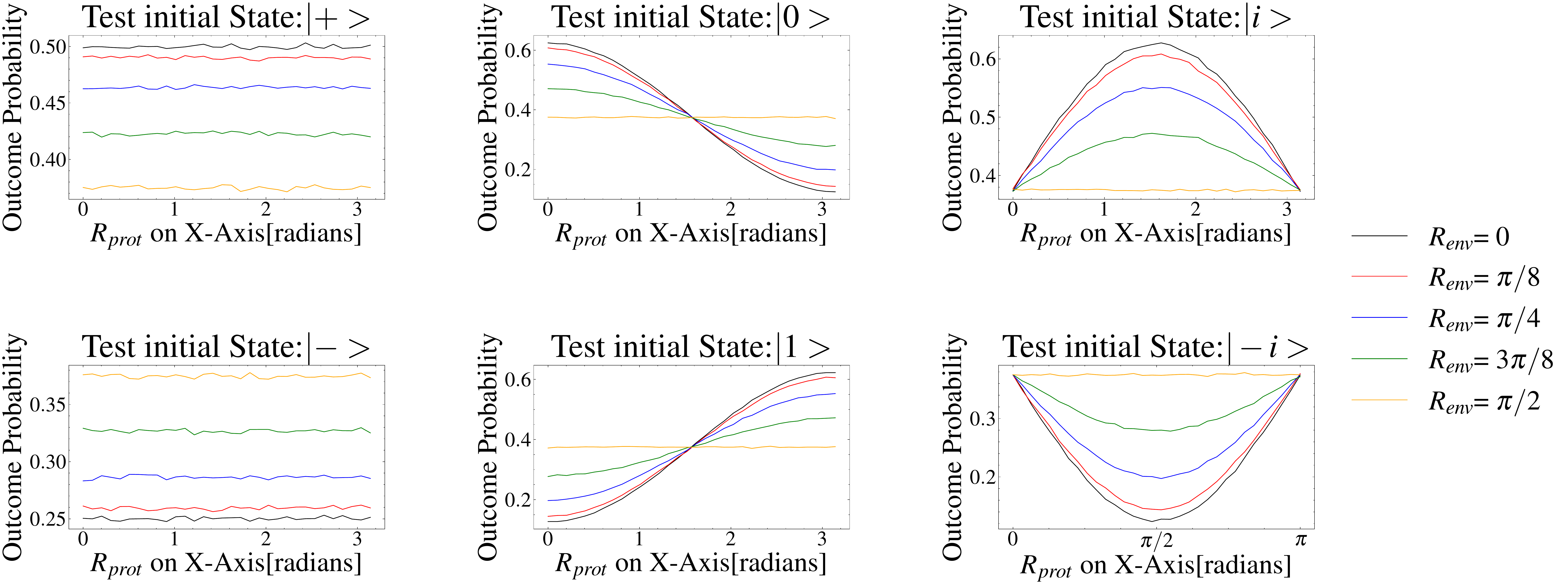}
\caption{Decoherence  Characterization  iteration  using  the  Toffoli  ST  with a \textit{control} qubit measurement, having the control qubit’s  state set to $\ket{+}$ and $R_{prot}$ set on the X-Axis.}
\label{fig:finalplottofoli}l
\end{figure}

This calculation result can be verified by the performed simulations, as presented in Figure \ref{fig:finalplottofoli}, where we also added the plots for all of the orthogonal states of the original ones. Note that states $\ket{+}$ and $\ket{-}$ lie along the X-axis, and therefore $R_{prot}(\alpha)$ has no effect on the outcome probability. In fact, a similar calculation shows that, for states alongside the X-axis, the probability of success for a modified Toffoli ST is given by:
\begin{equation*}
     Prob(0,0) = \frac{3}{8}\pm\frac{1}{8}\cos(\varepsilon),
\end{equation*}
where the sign is positive (negative) for states closer to $\ket{+}$ ($\ket{-}$).

\begin{figure}[ht]
\centering
  \includegraphics[width=1.0\linewidth]{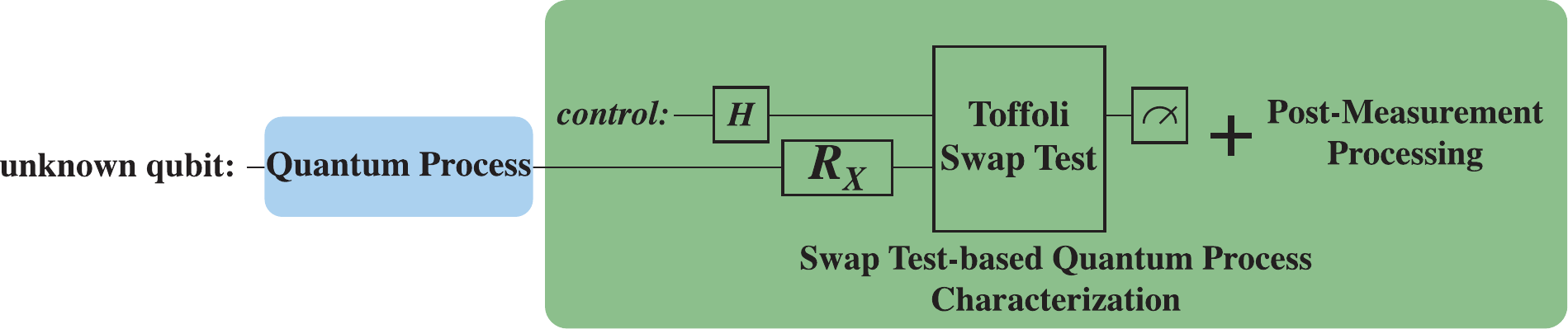}
\caption{Swap Test-based method for Quantum Process characterization.}
\label{fig:conclusionscheme}       % Give a unique label
\end{figure}

With that said, this protocol setting, presented in Figure \ref{fig:conclusionscheme} was able to both distinguish results for our test cases and demonstrate unique curves according to how the \textit{test} qubit was affected by decoherence. As we have tested for all of the Pauli Matrices eigenstates, this method is valid for any state on the Bloch Sphere. Interestingly enough, note that the method only employs unitary operations that correspond to rotations around the X-axis; in other words, we measure observables that correspond to axes belonging to a great circle on Bloch sphere that connects the poles and pass through the states $\ket{+i}$ and $\ket{-i}$. 

It should be noted that, according to Fig. \ref{fig:finalplottofoli} , as long as the input state is known except for its Bloch vector length, only one angle value is needed in order to find it out; except for angles 0, $\pi/2$ and $\pi$ (where the lines intersect in some cases), an analysis at any other angle would suffice, because a single point determines the full curve. This means that, in this particular scenario, no actual ``rotation" is needed, i.e., the method employs a static quantum circuit, in the same way as in a standard quantum state tomography.

\section{Comparing Swap Test Circuit Variants on real quantum computers}

Up to this point, we have employed Qiskit's Aer \textit{qasm\_simulator} backend \citet{cite_qiskit}; the next natural step is running a test bench for both ST variants on a real IBM quantum computer - \textit{ibmq\_manila} - in order to analyse the innate decoherence which is present in every quantum computer. This natural decoherence is due to the natural (unwanted) coupling between the qubits and the environment, which increases with the circuit depth.

In order to verify whether practical quantum computers have noticeable decoherence effects, even for short quantum circuits, we realized an experiment in which both \textit{test} and \textit{control} qubits were initially prepared at the $\ket{+}$ state, without any \aspas{artificial} decoherence process, with $R_{env}$ set to $0$. The \textit{test} qubit was gradually rotated along the Y-Axis by the $R_{prot}$ operation while the \textit{control} qubit remained unchanged. The \textit{test} qubit was rotated until the $\ket{-}$ state, where it would be orthogonal to the \textit{control} qubit's state.

\begin{figure}[ht]
\centering
\subfloat[Original Swap Test results]{\label{fig:originalstIBM}{\includegraphics[width=0.3\textwidth]{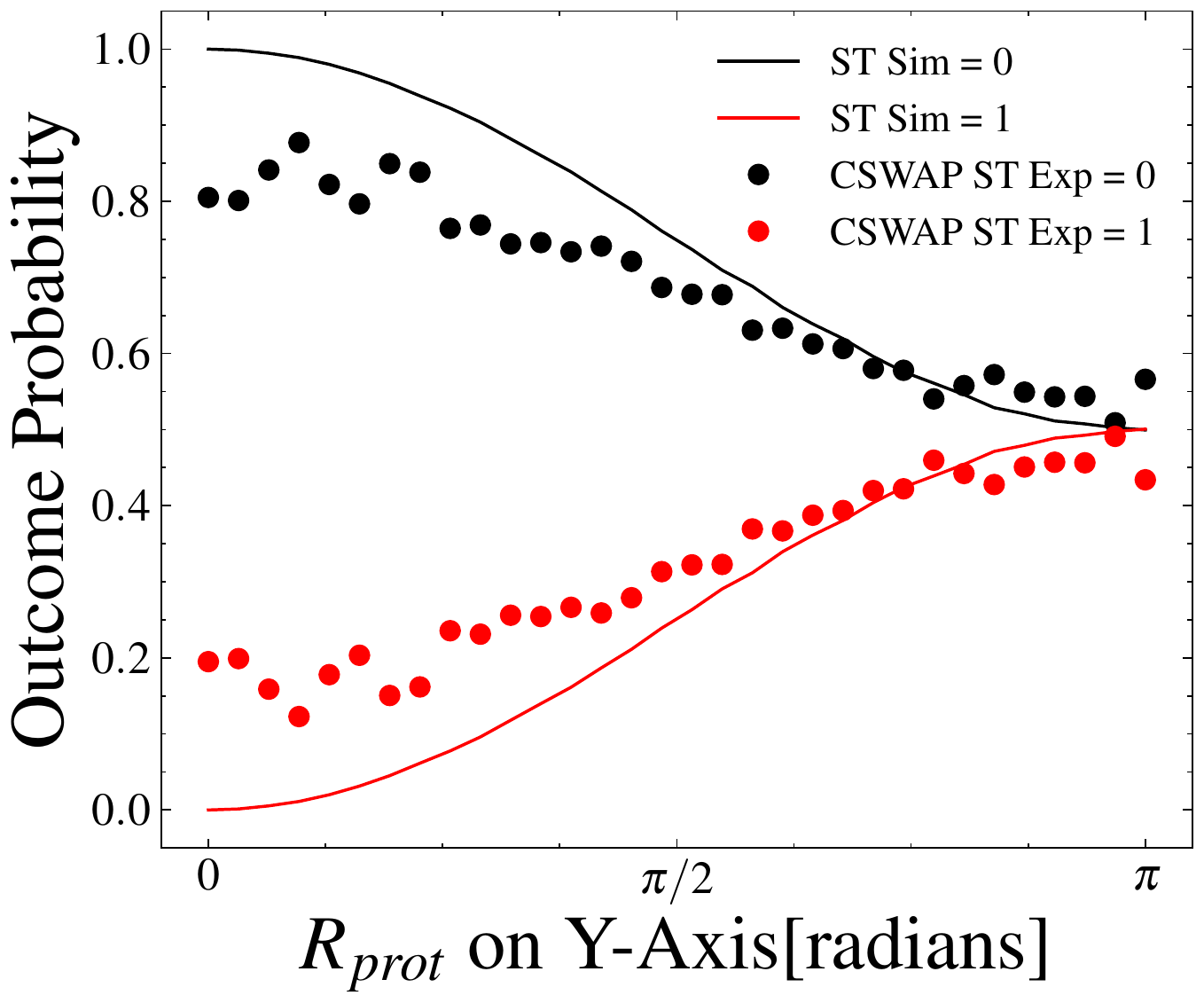}}}

\subfloat[BSM Swap Test results]{\label{fig:BSMstIBM}{\includegraphics[width=0.3\textwidth]{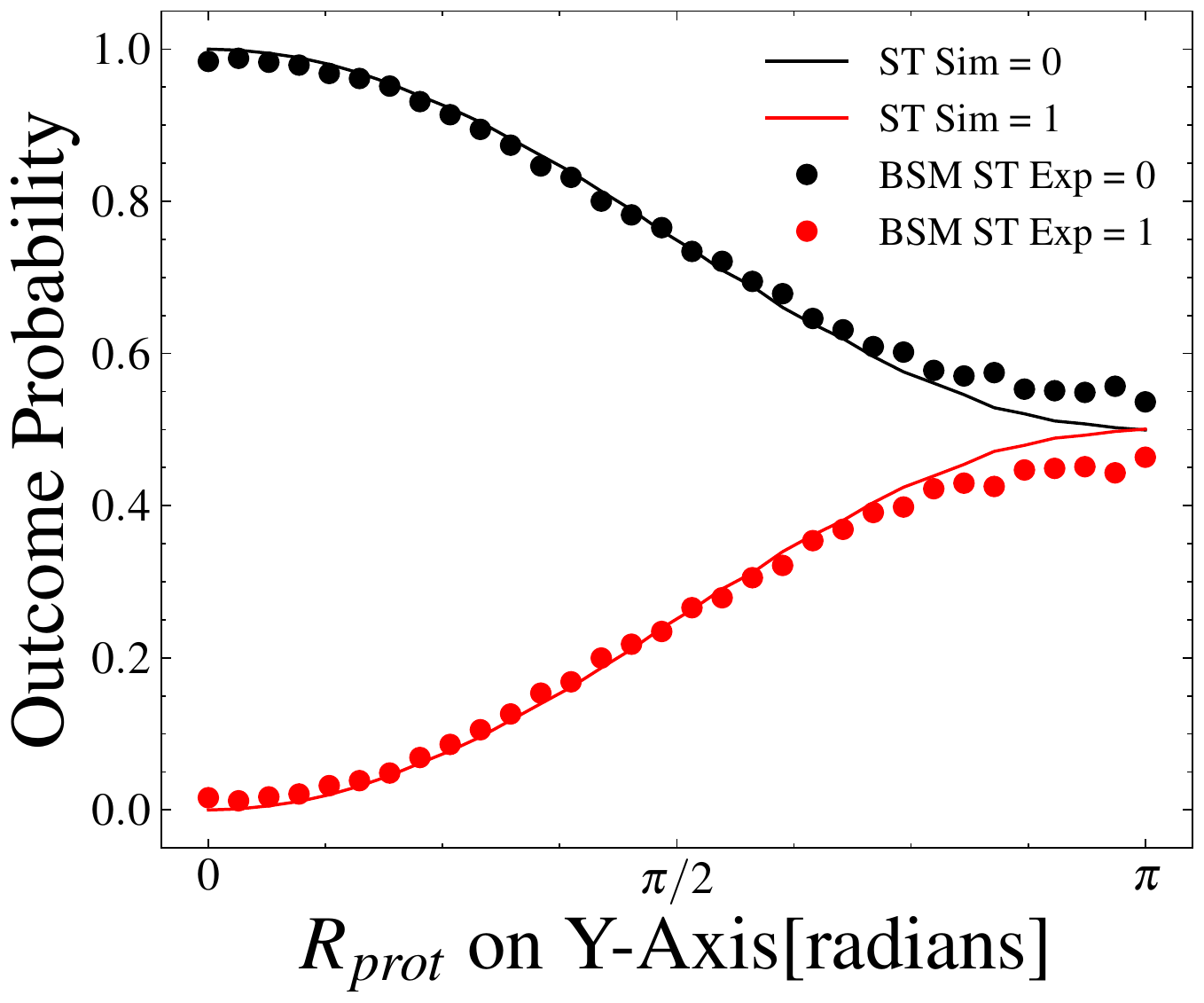}}}

\subfloat[Toffoli Swap Test results]{\label{fig:toffstIBM}{\includegraphics[width=0.3\textwidth]{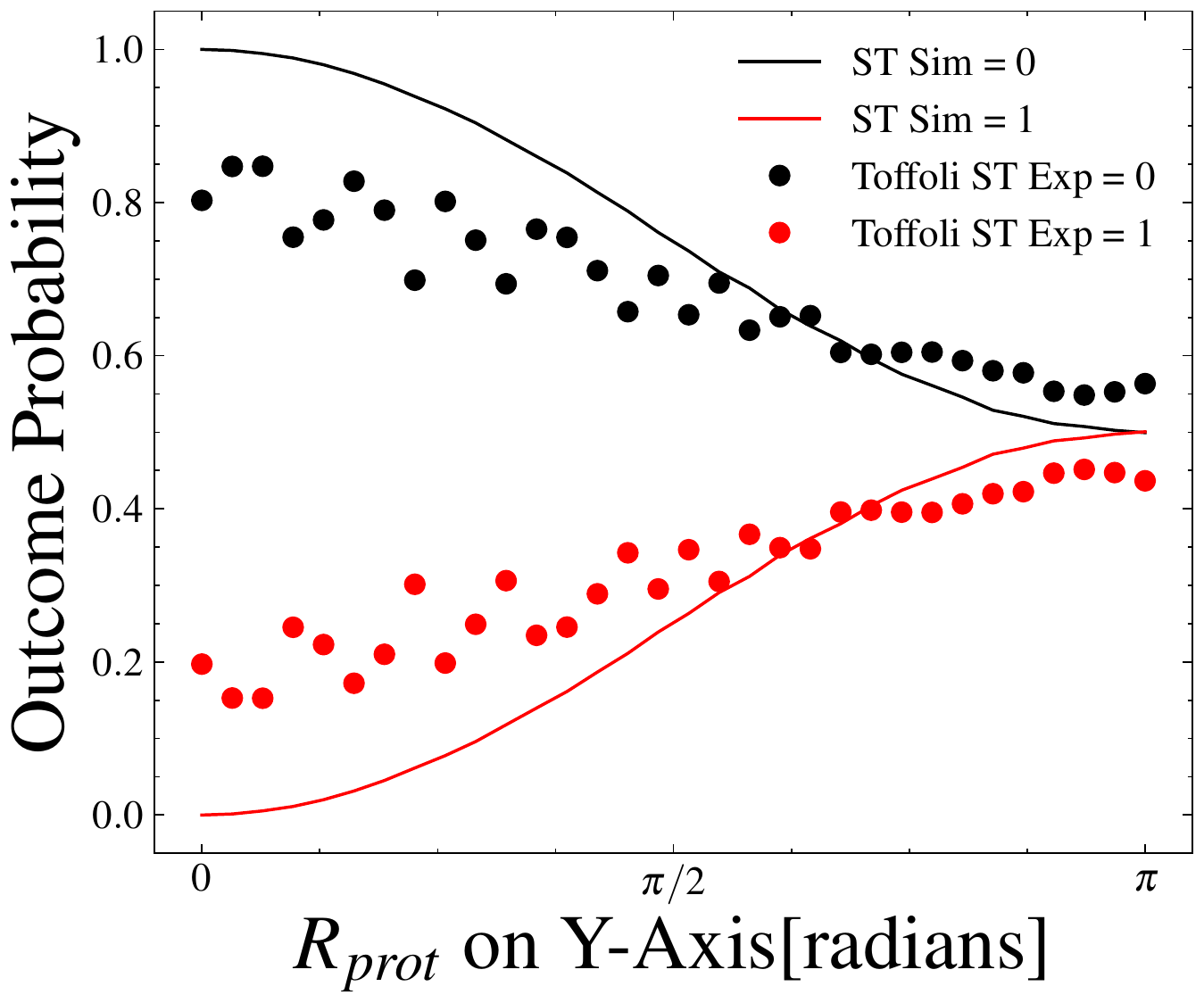}}}

% \begin{subfigure}{0.7\linewidth}
%   % \centering
%   \includegraphics[width=\linewidth]{simple_rot_ibm_original_st.pdf}
%   \caption{Original Swap Test results}
%   \label{fig:originalstIBM}
% \end{subfigure}%

% \begin{subfigure}{0.7\linewidth}
%   % \centering
%   \includegraphics[width=\linewidth]{simple_rot_ibm_bsm_st.pdf}
%   \caption{BSM Swap Test results}
%   \label{fig:BSMstIBM}
% \end{subfigure}%

% \begin{subfigure}{0.7\linewidth}
%   % \centering
%   \includegraphics[width=\linewidth]{simple_rot_ibm_toff_st.pdf}
%   \caption{Toffoli Swap Test results}
%   \label{fig:toffstIBM}
% \end{subfigure}%

\caption{Swap Test's outcomes on a real IBM quantum computer (\textit{ibmq\_manila}). The plots distinguish the simulation results (lines) from the actual experimental ones (dots). It can be seen that the BSM ST produces results that are much closer to the simulation results, when compared to the CSWAP ST and Toffoli Swap Tests.}

\end{figure}

Checking the outcomes from the original ST circuit - Figure \ref{fig:originalstIBM}, followed by the ones from the BSM ST circuit - Figure \ref{fig:BSMstIBM} - and the Toffoli ST circuit, it is possible to observe there is not any significant intrinsic decoherence in any of the ST circuits when executed by a real quantum computer. Consequently, we did not consider any real decoherence to our simulation results. However, it is clear that the BSM ST provides a better experimental output, as it is comprised of a shorter circuit depth and requires less qubits.

\section{Conclusion}
The Swap Test has been investigated as a tool for characterization of decoherence in quantum computers. We have shown that a simple comparison between a control and a test qubit using standard Swap Test schemes does not unveil enough information to determine with certainty whether the test qubit has suffered decoherence, i.e., if it's in a mixed state. A modified version of the Toffoli Swap Test, on the other hand, can be employed for successfully characterizing a decoherence process on a qubit, by performing the Swap Test between a known pure state (control) and a mixed state (test), where the latter is obtained by a controlled entanglement with an auxiliary qubit (environment). The process differs from standard quantum state tomography, but is able to obtain the same information (i.e., reconstruct the state's density matrix), with the possible advantage of not destroying the test qubit in the process, even though it is not kept unmodified in the case where decoherence takes place.

It should be mentioned that the full power of the Swap Test cannot be achieved unless we have access to internal degrees of freedom of the qubit, as happens for example in a Hong-Ou-Mandel interferometer with photonic qubits, which enables us to distinguish time-varying unitary operations from actual decoherence (i.e., entanglement with the environment). This examination will be performed in a future work.

\noindent \textbf{Funding.} The authors acknowledge the financial support of CAPES, CNPq and FAPERJ.

% Bibliography
\bibliography{main.bib}

\begin{thebibliography}{12}
\expandafter\ifx\csname natexlab\endcsname\relax\def\natexlab#1{#1}\fi
\expandafter\ifx\csname bibnamefont\endcsname\relax
  \def\bibnamefont#1{#1}\fi
\expandafter\ifx\csname bibfnamefont\endcsname\relax
  \def\bibfnamefont#1{#1}\fi
\expandafter\ifx\csname citenamefont\endcsname\relax
  \def\citenamefont#1{#1}\fi
\expandafter\ifx\csname url\endcsname\relax
  \def\url#1{\texttt{#1}}\fi
\expandafter\ifx\csname urlprefix\endcsname\relax\def\urlprefix{URL }\fi
\providecommand{\bibinfo}[2]{#2}
\providecommand{\eprint}[2][]{\url{#2}}

\bibitem[{\citenamefont{Preskill}(2018)}]{Preskill_2018}
\bibinfo{author}{\bibfnamefont{J.}~\bibnamefont{Preskill}},
  \bibinfo{journal}{Quantum} \textbf{\bibinfo{volume}{2}}, \bibinfo{pages}{79}
  (\bibinfo{year}{2018}), ISSN \bibinfo{issn}{2521-327X},
  \urlprefix\url{http://dx.doi.org/10.22331/q-2018-08-06-79}.

\bibitem[{\citenamefont{Helsen et~al.}(2020)\citenamefont{Helsen, Roth,
  Onorati, Werner, and Eisert}}]{helsen2020general}
\bibinfo{author}{\bibfnamefont{J.}~\bibnamefont{Helsen}},
  \bibinfo{author}{\bibfnamefont{I.}~\bibnamefont{Roth}},
  \bibinfo{author}{\bibfnamefont{E.}~\bibnamefont{Onorati}},
  \bibinfo{author}{\bibfnamefont{A.~H.} \bibnamefont{Werner}},
  \bibnamefont{and} \bibinfo{author}{\bibfnamefont{J.}~\bibnamefont{Eisert}},
  \emph{\bibinfo{title}{A general framework for randomized benchmarking}}
  (\bibinfo{year}{2020}), \eprint{2010.07974}.

\bibitem[{\citenamefont{Cross et~al.}(2019)\citenamefont{Cross, Bishop,
  Sheldon, Nation, and Gambetta}}]{PhysRevA.100.032328}
\bibinfo{author}{\bibfnamefont{A.~W.} \bibnamefont{Cross}},
  \bibinfo{author}{\bibfnamefont{L.~S.} \bibnamefont{Bishop}},
  \bibinfo{author}{\bibfnamefont{S.}~\bibnamefont{Sheldon}},
  \bibinfo{author}{\bibfnamefont{P.~D.} \bibnamefont{Nation}},
  \bibnamefont{and} \bibinfo{author}{\bibfnamefont{J.~M.}
  \bibnamefont{Gambetta}}, \bibinfo{journal}{Phys. Rev. A}
  \textbf{\bibinfo{volume}{100}}, \bibinfo{pages}{032328}
  (\bibinfo{year}{2019}),
  \urlprefix\url{https://link.aps.org/doi/10.1103/PhysRevA.100.032328}.

\bibitem[{\citenamefont{Amaral et~al.}(2018)\citenamefont{Amaral, Carneiro,
  Tempor{\~a}o, and von~der Weid}}]{amaral2018complementarity}
\bibinfo{author}{\bibfnamefont{G.~C.} \bibnamefont{Amaral}},
  \bibinfo{author}{\bibfnamefont{E.~F.} \bibnamefont{Carneiro}},
  \bibinfo{author}{\bibfnamefont{G.~P.} \bibnamefont{Tempor{\~a}o}},
  \bibnamefont{and} \bibinfo{author}{\bibfnamefont{J.~P.} \bibnamefont{von~der
  Weid}}, \bibinfo{journal}{JOSA B} \textbf{\bibinfo{volume}{35}},
  \bibinfo{pages}{601} (\bibinfo{year}{2018}).

\bibitem[{\citenamefont{Hong et~al.}(1987)\citenamefont{Hong, Ou, and
  Mandel}}]{PhysRevLett.59.2044}
\bibinfo{author}{\bibfnamefont{C.~K.} \bibnamefont{Hong}},
  \bibinfo{author}{\bibfnamefont{Z.~Y.} \bibnamefont{Ou}}, \bibnamefont{and}
  \bibinfo{author}{\bibfnamefont{L.}~\bibnamefont{Mandel}},
  \bibinfo{journal}{Phys. Rev. Lett.} \textbf{\bibinfo{volume}{59}},
  \bibinfo{pages}{2044} (\bibinfo{year}{1987}),
  \urlprefix\url{https://link.aps.org/doi/10.1103/PhysRevLett.59.2044}.

\bibitem[{\citenamefont{Buhrman et~al.}(2001)\citenamefont{Buhrman, Cleve,
  Watrous, and de~Wolf}}]{Buhrman_2001}
\bibinfo{author}{\bibfnamefont{H.}~\bibnamefont{Buhrman}},
  \bibinfo{author}{\bibfnamefont{R.}~\bibnamefont{Cleve}},
  \bibinfo{author}{\bibfnamefont{J.}~\bibnamefont{Watrous}}, \bibnamefont{and}
  \bibinfo{author}{\bibfnamefont{R.}~\bibnamefont{de~Wolf}},
  \bibinfo{journal}{Physical Review Letters} \textbf{\bibinfo{volume}{87}}
  (\bibinfo{year}{2001}), ISSN \bibinfo{issn}{1079-7114},
  \urlprefix\url{http://dx.doi.org/10.1103/PhysRevLett.87.167902}.

\bibitem[{\citenamefont{Garcia-Escartin and
  Chamorro-Posada}(2013)}]{garcia2013swap}
\bibinfo{author}{\bibfnamefont{J.~C.} \bibnamefont{Garcia-Escartin}}
  \bibnamefont{and}
  \bibinfo{author}{\bibfnamefont{P.}~\bibnamefont{Chamorro-Posada}},
  \bibinfo{journal}{Physical Review A} \textbf{\bibinfo{volume}{87}},
  \bibinfo{pages}{052330} (\bibinfo{year}{2013}).

\bibitem[{\citenamefont{Cincio et~al.}(2018)\citenamefont{Cincio,
  Suba{\c{s}}{\i}, Sornborger, and Coles}}]{Cincio_2018}
\bibinfo{author}{\bibfnamefont{L.}~\bibnamefont{Cincio}},
  \bibinfo{author}{\bibfnamefont{Y.}~\bibnamefont{Suba{\c{s}}{\i}}},
  \bibinfo{author}{\bibfnamefont{A.~T.} \bibnamefont{Sornborger}},
  \bibnamefont{and} \bibinfo{author}{\bibfnamefont{P.~J.} \bibnamefont{Coles}},
  \bibinfo{journal}{New Journal of Physics} \textbf{\bibinfo{volume}{20}},
  \bibinfo{pages}{113022} (\bibinfo{year}{2018}),
  \urlprefix\url{https://doi.org/10.1088/1367-2630/aae94a}.

\bibitem[{\citenamefont{Foulds et~al.}(2021)\citenamefont{Foulds, Kendon, and
  Spiller}}]{foulds2021controlled}
\bibinfo{author}{\bibfnamefont{S.}~\bibnamefont{Foulds}},
  \bibinfo{author}{\bibfnamefont{V.}~\bibnamefont{Kendon}}, \bibnamefont{and}
  \bibinfo{author}{\bibfnamefont{T.}~\bibnamefont{Spiller}},
  \emph{\bibinfo{title}{The controlled swap test for determining quantum
  entanglement}} (\bibinfo{year}{2021}), \eprint{2009.07613}.

\bibitem[{\citenamefont{\textit{et al}}(2019)}]{cite_qiskit}
\bibinfo{author}{\bibfnamefont{H.~A.} \bibnamefont{\textit{et al}}},
  \emph{\bibinfo{title}{Qiskit: An open-source framework for quantum
  computing}} (\bibinfo{year}{2019}).

\bibitem[{\citenamefont{Peres}(1995)}]{aperes}
\bibinfo{author}{\bibfnamefont{A.}~\bibnamefont{Peres}},
  \emph{\bibinfo{title}{Quantum Theory: Concepts and Methods}}
  (\bibinfo{publisher}{Kluwer Academic Publishers},
  \bibinfo{address}{Netherlands}, \bibinfo{year}{1995}), ISBN
  \bibinfo{isbn}{0792325494}.

\bibitem[{\citenamefont{Nielsen and Chuang}(2011)}]{10.5555/1972505}
\bibinfo{author}{\bibfnamefont{M.~A.} \bibnamefont{Nielsen}} \bibnamefont{and}
  \bibinfo{author}{\bibfnamefont{I.~L.} \bibnamefont{Chuang}},
  \emph{\bibinfo{title}{Quantum Computation and Quantum Information: 10th
  Anniversary Edition}} (\bibinfo{publisher}{Cambridge University Press},
  \bibinfo{address}{USA}, \bibinfo{year}{2011}), \bibinfo{edition}{10th} ed.,
  ISBN \bibinfo{isbn}{1107002176}.

\end{thebibliography}

\end{document}